\documentclass[10pt,a4paper]{article}
\usepackage{moreverb}
\usepackage{float}

\usepackage[dvipdfmx]{graphicx}  
\usepackage{ascmac}
\usepackage{amsmath}
 
\usepackage{array}   

\begin{document}
\title{Zipf's law for share price and company fundamentals}
\author {Taisei Kaizoji and Michiko Miyano \\ Graduate School of Arts and Sciences,\\ International Christian Unversity}
\maketitle
\begin{center}
abstract
\end{center}
We investigate statistically the distribution of share price and the distributions of three common financial indicators using data from approximately 8,000 companies publicly listed worldwide for the period 2004-2013. We find that the distribution of share price follows Zipf's law; that is, it can be approximated by a power law distribution with exponent equal to 1. An examination of the distributions of dividends per share, cash flow per share, and book value per share - three financial indicators that can be assumed to influence corporate value (i.e. share price) - shows that these distributions can also be approximated by a power law distribution with power-law exponent equal to 1. We estimate a panel regression model in which share price is the dependent variable and the three financial indicators are explanatory variables. The two-way fixed effects model that was selected as the best model has quite high power for explaining the actual data. From these results, we can surmise that the reason why share price follows Zipf's law is that corporate value, i.e. company fundamentals, follows Zipf's law.

\section{Introduction}
Zipf's law is an empirical law indicating a relationhip between the frequency of an event and its rank. It is often formulated by using the complementary cumulative distribution function following a power law,  $Pr(X>x)=cx^{-\alpha}$ where $Pr(X>x)$  denotes the probability that a stochastic variable, $X$, is larger than  $x$ , $c$   is a constant and  is the power-law exponent. Zipf's law states that the power exponent, $\alpha$ , is equal to unity. 

In his book, published in 1932 (Zipf 1932), G. K. Zipf reported that the number of times that all the various words appear in human language closely follows this law\footnote{Pareto(1897) before Zipf indicated that the distribution of individual incomes follows a power law distribution.} : Zipf's law has since been found to apply in a variety of fields ; Zipf's law for human language (Miller (1958)); Zipf's law for monkey-typing texts (Li (1992)), Zipf's law for the number of people living in a city (Hill 1970, Ijiri and Simon (1977), Gabaix 1999), Zipf's law for the number of website visits (Cunha, C.R., Bestavros, A., Crovella, M.E. (1995)). Zipf's law for earthquakes (Sornette, D., Knopoff, L., Kagan, Y.Y., Vanneste, C. (1996)), Zipf's law for US company sizes (Axtell (2001)), Zipf's law for  human transcriptomes (Ogasawara, S. Kawamoto, and K. Okubo (2003)), Zipf's law for gene expression (Furusawa and K. Kaneko (2003)), and the others. 

For several decades, a number of theoretical studies to explain power-law laws and Zipf's law have been presented by researchers working in various academic fields. For example, Simon (1955), Mandelbrot (1961), Montroll and Shlesinger (1982), Angle (2006), Zanette and Manrubia (1997), Marsili, and Zhang (1998), Gabaix (1999), Reed (2000), Chakrabarti and Chakrabarti (2010), and Irie and Tokita (2012), and others.

This paper investigates Zipf's law for share prices and the reason why share prices follow Zipf's law. Although there are a number of empirical studies on Zipf's law, little attention has been given to examining Zipf's law in terms of statistical significance testing. This paper takes on precisely this task. First, using 7,796 companies listed in world-wide stock markets, we demonstrate that share prices follow Zipf's law. Secondly, we introduce an empirical model proposed by Kaizoji and Miyno (2016b) to estimate company fundamentals for these 7,796 companies. The empirical model is a panel regression model (a two-way fixed effects model) that calculates company fundamentals by using dividends per share, cash flow per share, and book value per share as explanatory variables. Thirdly, we investigate whether the distribution of company fundamentals follows Zipf's law. We find that each of the three financial indicators follows Zipf's law and that the distribution of fundamentals matches substantially the distribution of share prices. On these grounds, we conclude that Zipf's law for stock price is caused by Zipf's law for company fundamentals.

The remainder of this paper is organized as follows: Section 2 explains the data set used; Section 3 demonstrates Zipf's law for share prices; Section 4 introduces an empirical model proposed by Kaizoji and Miyano (2016b) to estimate company fundamentals; Section 5 establishes the applicability of Zipf's law for company fundamentals; Section 6 shows Zipf's law for the three financial indicators - dividends per share, cash flow per share, and book value per share - that serve as the explanatory variables in the empirical model introduced in Section 4; Section 7 proposes a stochastic model exploring Zipf's law for company fundamentals; Section 8 offers concluding remarks.

\section{Data}
The data used in this paper come from the OSIRIS database provided by Bureau Van Dijk. The database contains financial information on globally listed industrial companies. We use annual data for the period 2004-2013. After extracting financial data and stock data for 7,796 companies over a 10-year period, we performed a statistical investigation of share price (at closing date) and three financial indicators. The financial indicators - dividends per share, cash flow per share, and book value per share - were obtained by dividing the values provided in the database by the number of shares outstanding. 
 
\section{Zipf'slaw for share price}   
Figure 1 shows the complementary cumulative distribution of share prices plotted with logarithmic horizontal and vertical axes. Share prices in this distribution are pooled from the 7,796 companies included in the study for the period 2004-2013. In all, there are 47,161 total observations. The right tail of this distribution is found to be approximately linear. That is, the distribution can be approximated by a power law distribution, written as 
\begin{equation}
  Pr(X > x) =cx^{-\alpha}   
\end{equation}
where $c$  is a constant and $\alpha$ denotes the power-law exponent. 

Figure 1: Zipf's law for share price: The complementary cumulative distribution of share prices (log-log plot) based on 47,161 share price observations pooled from 7,796 companies for the period 2004-2013. The power-law exponent, estimated using the MLE (maximum likelihood estimator) method, is 1.003.

As a first step, we estimated the power-law exponent, Although several estimators can be used to estimate the value of the power-law exponent, this paper uses the maximum likelihood estimator (MLE), which is commonly applied. The maximum likelihood estimator of exponent $\alpha$  is \footnote{Details of the derivations are given in Appendix}

\begin{equation}
\hat{\alpha}=n[\sum_{i=1}^{n}\ln (\frac{x_{i}}{x_{min}}) ]^{-1}
\end{equation}

where, $x_{i}, i=1,\cdots , n$  are the observed values of   such that $x_{i}\ge x_{min}$  , and  $\hat{\alpha}$ denotes the estimate of $\alpha$. Using the share price data pooled from 7,796 companies for the period 2004-2013, the estimated power-law exponent is very close to unity, $\hat{\alpha}=1.003$ . The results are presented in the first column of Table 1. 

We then test whether that complementary cumulative distribution of share prices follows a power law distribution by using a goodness-of-fit test based on a measurement of distance between the empirical  distribution function and the fitted   distribution function. The distance is commonly measured either by a supremum norm or a quadratic norm. The most well-known supremum statistic is the Kolmogorov-Smirnov statistic written as

\begin{equation}
 D=\sup_{x} |F_{n}(x)-F(x)|     
\end{equation} 
On the other hand, the Cramer-von Mises family statistic in a quadratic norm is written as  
\begin{equation}
Q=n\int_{-\infty}^{\infty}[F_{n}(x)-F(x)]^{2}\psi (x)dF(x) 
\end{equation} 
When $\psi(x)=1$ ,   is the Cramer-von Mises statistic, denoted $W^{2}$  . Although a number of other of goodness-of-fit statistics have been proposed, we use the   statistic of Cramer-von Mises in this study\footnote{Clauset, et. al. (2009) found that Anderson and Darling statistics are highly conservative in their application, reducing the ability to validate the power law model unless there are many samples in the tail of distribution.}.  The null hypothesis that the empirical distribution function is a power law distribution is rejected when a small $p$-value is obtained. In our study, the null hypothesis is not rejected at the 10\% significant level since we obtain a $p$-value of 0.132. That is, the Cramer-von Mises test statistic indicates that the distribution of share prices follows a power law. 

In the second step, we verify that share price follows Zipf's law using two test procedures. One is the Lagrange Multiplier (LMZ) test proposed by Urzua (2000); the other is the Likelihood Ratio (LR) test. The Likelihood Ratio test can be used in the case of large samples, while the Urzua test can be applied to cases involving small samples\footnote{Urzua(2000) uses the table for significanct points for LMZ. The table is presented in the Apendix.} . Test results are shown in Table 1. In both tests, the null hypothesis that the power-law exponent is equal to 1 cannot be rejected at the 10\% significance level since the Chi-squared test statistics are sufficiently small.

In summary, we can confirm that the distribution of share price follows Zipf's law in the right tail, which includes 2\% of the total observations. 

\begin{table}[H]
\begin{center}
\begin{tabular}{ccccccc} \hline
power-law & $X_{min}$ & Cramer-von& Lagrange &Likelihood & Tail\\
exponent              &        &  Mises test& Multiplier & Ratio &\\ 
     &    &  $p$-value &  LMZ & LR & n \\ \hline 
1.003 & 133.4 &0.132 & 0.013 & 0.010 & 943 \\ \hline
\end{tabular}
\end{center} 
\caption{Test for Zipf's law for share price: Estimation of the power-law exponent of share price, and test results for the power law distribution and for Zipf's law. LMZ and LR test involve test-statistics of the Chi-squared distribution. The null hypothesis that the power law exponent is equal to 1 cannot be rejected at the 10\% significance level.}
\end{table}

\section{Econometric model for share price(Kaizoji and Miyano(2016b))}
In the previous section, it was shown that Zipf's law holds for share prices. This section investigates the relationship between share price and company fundamentals. To that end, we use the panel regression model proposed by Kaizoji and Miyano (2016b) to estimate company fundamentals. Since the database used in this research contains cross-sectional data for the period 2004-2013, panel analysis is appropriate. The model formulates the relationship between share prices and three financial indicators - dividends per share, cash flow per share, and book value per share - commonly used in models that evaluate corporate valuation, so-called company fundamentals. The econometric model can be formally written as 
\begin{equation}
 \ln Y_{it} = a + b_{1}\ln X_{1,it} + b_{2}\ln X_{2,it} + b_{3} \ln X_{3,it} + u_{it}  \quad i=1,\cdots, N;  t=1,\cdots, T
\end{equation} 

where $Y_{it}$ denotes the dependent variable (share price) for company $i$   in year $t$ ;  $a$ denotes a constant; $X_{1,it}$ is the dividends per share of company $i$   in year $t$ ; $X_{2,it}$  is the cash flow per share of company $i$  in year $t$  ; $X_{3,it}$  is the book value per share of company $i$  in year $t$ ; $u_{it}$  denotes the error term. 

We estimate the model in equation (5) using the Panel Least Squares method. In the panel regression model, the error term,  $u_{it}$   , can be assumed to be divided into a pure disturbance term and an error term due to other factors. Assuming a two-way error component model with respect to error, the factors other than disturbance are (i) factors due to unobservable individual effects, and (ii) factors due to unobservable time effects. That is, the error term can be written as

\begin{equation}
 u_{it} =\mu_{i} + \gamma_{t} + \epsilon_{it}    
\end{equation} 

where $\mu_{i}$   denotes unobservable individual effects,  $\gamma_{t}$ denotes unobservable time effects, and  $\epsilon_{it}$ denotes pure disturbance. 

If both  $\mu_{i}$    and   $\gamma_{t}$  are equal to zero, equation (5) is estimated using the pooled OLS method. If either      $\mu_{i}$   or $\gamma_{t}$    is equal to zero, equation (6) is a one-way error component model. If both $\mu_{i}$   and    $\gamma_{t}$  are not equal to zero, equation (6) is a two-way error component model. There are two estimation methods for estimating the error term in equation (6). One is fixed effects estimation and the other is random effects estimation. Therefore, the available estimation models are a pooled OLS, an individual fixed effects model, a time effects model, a two-way fixed effects model, an individual random effects model, a time random effects model, and a two-way random effects model\footnote{We used the EViews software package to estimate the models. The two-way random effects model was unavailable since we used unbalanced panel data.} .

We estimated the models described above and found, after appropriate model selection tests, that the two-way fixed effects model was the best model. The model selection tests used in this study include the likelihood ratio test and F-test for the selection of the pooled OLS model vs the fixed effects model, and the Hausman test for the selection of the random effects model vs the fixed effects model. The selection test for the pooled OLS model vs the random effects model is based on the simple test proposed by Wooldridge (2010)\footnote{Wooldridge (2010, p.299) proposed the method that uses residuals from pooled OLS to check the existence of serial correlation.} 
  The two-way fixed effects mode is written as
\begin{gather}
 \ln Y_{it} = a + b_{1}\ln X_{1,it} + b_{2}\ln X_{2,it} + b_{3} \ln X_{3,it} + \epsilon_{it}  \\
a=a_{0} + \mu_{i} + \gamma_{t}  \notag
\end{gather}
where  $a_{0}$  is a constant term common to all companies,  $\mu_{i}$  denotes the individual fixed effects, and  $\gamma_{t}$   denotes the time fixed effects.   $\mu_{i}$  is constant toward time series and $\gamma_{t}$  is constant toward cross section. $\epsilon_{it}$  is the pure disturbance. The individual fixed effects,  $\mu_{i}$   , account for an individual company's heterogeneity and includes such factors as the company's diversity of corporate governance and the quality of its employees.  The time fixed effects,$\gamma_{t}$ , indicate variables that fluctuate over time but are fixed across companies. The time fixed effects reflect various shocks, including financial shocks. 
  
Table 2 shows the results of the Kaizoji and Miyano (2016b) panel regression model described in equation (7). The signs of the three coefficients are all positive, which is consistent with corporate value theory. The $p$-values for the coefficients are quite small, indicating statistical significance in all three cases. In addition, the R-squared value is 0.97, indicating that the estimated model explains much of the variation in share prices.

\begin{table}[H]
\begin{center}
\begin{tabular}{lrrrrr} \hline
   & a0 & b1 & b2 & b3 &$R^{2}$ \\ \hline 
Coefficient&1.485 & 0.137 &0.298 &0.378& 0.969 \\ 
Std. error & 0.014 & 0.003 & 0.004 & 0.007&  \\ 
$p$-value  &  0.000 & 0.000 & 0.000 & 0.000 & \\ \hline
\end{tabular}
\end{center} 
\caption{Results of the Panel Regression model (two-way fixed effects model) provided by Kaizoji and Miyano (2016b). Total panel (unbalanced) observations are 47,161. The $p$-values for the coefficients indicate statistical significance in all cases. The R-squared value is 0.97.}
\end{table}

The model (7) is found to have quite high explanatory power with respect to share price, as the results shown above indicate. From a business management point of view, companies can maximize their share price by enhancing dividends per share, cash flow per share, and book value per share. 

Estimates of the two-way fixed effects model for share price, $\ln \hat{Y}_{it}$  , can be shown as 
\begin{equation}
\ln \hat{Y}_{it} = \hat{a_{0}} + \hat{\mu_{i}} + \hat{\gamma_{t}} + \hat{b_{1}}\ln X_{1,it} + \hat{b_{2}}\ln X_{2,it} + \hat{b_{3} }\ln X_{3,it}
\end{equation} 
We call $\hat{Y}_{it}$  the theoretical share price.
\section{Zipf's law for company fundamentals}

In the previous section, we found that the theoretical share price as estimated using a two-way fixed effects model fits actual share price very well, indicating that the two-way fixed effects model explains actual share price quite well. However, this result does not explain the reason why the distribution of share price follows Zipf's law. This section examines company fundamentals, defined as the optimal share price reflecting corporate value. 

Kaizoji and Miyano (2016b) proposed computing company fundamentals by eliminating the time fixed effects term,  , from the theoretical share price described in equation (8), 

\begin{equation}
\ln \tilde{Y}_{it} = \hat{a_{0}} + \hat{\mu_{i}} + \hat{b_{1}}\ln X_{1,it} + \hat{b_{2}}\ln X_{2,it} + \hat{b_{3} }\ln X_{3,it}
\end{equation} 
where $\tilde{Y}_{it}$  denotes the fundamentals of company $i$   in year $t$  . 

Figure 2 shows the complementary cumulative distributions of actual share price and company fundamentals plotted with logarithmic horizontal and vertical axes. At a glance, Figure 2 suggests that the distribution of fundamentals matches substantially the distribution of actual share prices. Indeed, the two-sample Kolmogorov-Simirnov test leads us to accept the null hypothesis affirming the coincidence of actual share price and company fundamentals (K-statistic = 0.007; $p$-value = 0.253). In short, the complementary cumulative distribution of share price coincides with that of company fundamentals.
 
We next estimate the power-law exponents using the MLE (maximum likelihood estimator) method. Comparing the power-law exponent of company fundamentals with that of actual share price and that of the theoretical share price, we found that the three power-law exponents are all close to unity: 1.003 for the actual share price, 1.012 for the theoretical share price, 1.006 for the fundamentals.

We then use the Cramer-von-Mises test to determine whether the complementary cumulative distributions of share prices, theoretical share prices as described in equation (8), and company fundamentals as described in equation (9) are power law distributions. Large $p$-values in each of the tests indicate that, for each of the three cases, the null hypothesis affirming that the complementary cumulative distribution is a power law distribution cannot be rejected. (The $p$-values were 0.132 (actual share price), 0.106 (theoretical share price), and 0.128 (company fundamentals).) 

Finally, we investigate whether company fundamentals and theoretical share price follow Zipf's law. In both cases, both the Urzua (2000) test and the Likelihood Ratio test do not reject, at the 10\% significance level, the null hypothesis affirming that the distribution follows Zipf's law (as was also true for actual share price). In addition, we found that the complementary cumulative distribution of the theoretical share price and that of company fundamentals follow Zipf's law in the right tail, which includes 2\% of the total observations.

Table 3 summarizes the results of estimating power-law exponents and the results of tests for power law and Zipf's law for actual share prices,   theoretical share prices, and company fundamentals. The data are pooled from 7,796 companies for the period 2004-2013; total observations are 47,161.

\begin{table}[H]
\begin{center}
\begin{tabular}{lccccccc} \hline
Share price &power-law & $X_{min}$ & Cramer-von-&  Lagrange &Likelihood & Tail\\
              & exponents   &    &  Mises test&  Multiplier& Ratio& \\ 
              &    &    & $p$-value &  LMZ &  LR & n  \\ \hline
actual  & 1.003 & 133.4 & 0.132 & 0.013 & 0.010 & 943 \\
theoretical  & 1.012 & 128.5 & 0.106 & 0.187 & 0.136 & 943 \\
fundamental&1.006 & 119.7 &0.128 & 0.032 & 0.032 & 1,000 \\ \hline
\end{tabular}
\end{center} 
\caption{Test for Zipf's law for share price: LMZ and LR are test-statistics of the Chi-squared distribution. The null hypothesis affirming that the power law exponent is equal to 1 cannot be rejected at the 10 \% significance level.}
\end{table}

In summary, we can confirm that the distributions of share price and company fundamentals follow Zipf's law in the right tail, which includes 2\% of total observations.

From the results described above, it is verified that the complementary cumulative distribution of fundamentals obtained from equation (9) closely coincides with the complementary cumulative distribution of actual share price. Therefore, we can infer that Zipf's law for share price is caused by Zipf's law for company fundamentals. 

Figure 2: The complementary cumulative distribution of actual share price and company fundamentals (log-log plot). Black indicates actual share price, red indicates company fundamentals. The complementary cumulative distribution of fundamentals coincides statistically with that of actual share price.

\section{Zipf's law for financial indicators per share}

In the previous section, we showed that the distribution of company fundamentals, estimated using a two-way fixed effects model, coincides with the distribution of actual share price. This result suggests that company fundamentals significantly affect actual share price. However, it does not explain the reason why company fundamentals follows Zipf's law. To consider the reason why Zipf's law for company fundamentals holds, we statistically investigate the distributions of the explanatory variables in the two-way fixed effects model described in equation (7).
 
Figures 3 through 5 show the plots of the complementary cumulative distributions for the three per share financial indicators using pooled data for the period 2004-2013. Figure 3 shows the distribution of dividends per share, Figure 4 shows the distribution of cash flow per share, and Figure 5 shows the distribution of book value per share. It is obvious that the right tails of the three distributions can be approximately linearized. 

Following the same procedure that was used for share price and company fundamentals, we estimated the power-law exponent for each of the three cases, producing estimates of 1.015 for dividend per share, 1.051 for cash flow per share, and 0.955 for book value per share. For all three distributions, Cramer-von-Mises tests indicate that the null hypothesis affirming that these are power-law distributions cannot be rejected ($p$-values are, 0.130, 0.151, and 0.221, respectively). Moreover, results from applying Urzua's test (LMZ) and the Likelihood Ratio (LR) test indicate that, in all three cases, the null hypothesis affirming power-law exponents cannot be rejected at the 10\% significance level. Table 4 presents the results. In summary, all three financial indicators per share follow Zipf's law.
 
From the results described here, we can infer that the reason Zipf's law holds for company fundamentals is that the distributions of the three financial indicators per share, representing corporate value, each follow the power law distribution with a power-law exponent equal to 1. Thus, we can conclude that the reason Zipf's law holds for company fundamentals is that Zipf's law holds for dividends per share, cash flow per share, and book value per share. 

Figure 3: Zipf's law for dividends per share. The complementary cumulative distribution of dividends per share (log-log plot) using 47,161 observations based on pooled data from 7,796 companies for the period 2004-2013. The power-law exponent estimated by the MLE (maximum likelihood estimator) method is 1.015.
 
Figure 4: Zipf's law for cash flow per share. The complementary cumulative distribution of cash flow per share (log-log plot) using 47,161 share price observations based on pooled data from 7,796 companies for the period 2004-2013. The power-law exponent estimated by the MLE (maximum likelihood estimator) method is 1.051.

Figure 5: Zipf's law for book value per share. The complementary cumulative distribution of the book value per share (log-log plot) using 47,161 share price observations based on pooled data from 7,796 companies for the period 2004-2013. The power-law exponent estimated by the MLE (maximum likelihood estimator) method is 0.955. 

\begin{table}[H]
\begin{center}
\begin{tabular}{lccccccc} \hline
financial  indicator & power-law  & $X_{min}$ & Cramer-von-& Lagrange&Likelihood & Tail\\
per share     & exponents     &        &  Mises test& Multiplier& Ratio& \\  
      &        &        &  $p$-value   &LMZ  &  LR  & n \\ \hline 
Dividends  & 1.015 & 3.6 &0.130 & 0.946 & 0.292 & 1250 \\ 
Cash Flow & 1.051 & 21.9 & 0.151 &  3.809 & 2.322 & 943 \\ 
Book Value & 0.955 & 98.9 & 0.221 &  2.828 & 2.010 & 943 \\ \hline
\end{tabular}
\end{center} 
\caption{Tests for Zipf's law for the three financial indicators. LMZ and LR are test-statistics of the Chi-squared distribution. The null hypothesis affirming that the power-law exponent is equal to 1 cannot be rejected at the 10\% significance level.}
\end{table}

\section{A stochastic model of company fundamentals}

In Section 3, we showed that our measure of company fundamentals follows Zipf's law. This means that there exists an extreme disparity in company fundamentals. In this section, we propose a model to explain the extreme disparity in company fundamentals from a theoretical point of view. It is easy to imagine that there are various factors explaining company fundamentals. For example, the quality of company employees, the foresight of company executives, production technology, and corporate governance can be listed as contributing factors. The formation of corporate value has its origin in the compound multiplier effects of these factors. Thus, the disparity in company fundamentals can be considered a reflection of differences in those factors, their number and influence, that affect corporate value. It can be assumed that there are a relatively few companies that have an abundance of strongly positive factors, while a great many companies have few such positive factors.
 
We propose a simple model which formalizes this notion\footnote{The model proposed here is mathematically the same as the model of income distributions proposed by Reed (2004)}.  Consider a variable,$Z_{n}$  , which describes a factor that affects the fundamentals of a company. 

If  $Z_{n}$  is greater than 1, the factor represented will enhance company fundamentals. Inversely, if $Z_{n}<1$ , then the corresponding factor will have a negative effect and reduce company fundamentals. We assume the following: 

\textbf{Assumption 1 (Compound multiplier effects):} $Z_{n}$  represents a set of stochastic variables that follow identical independent distributions. We can assume that company fundamentals can be determined by multiplying the $Z_{n}$ variables. That is, the value of company fundamentals is defined by the equation

\begin{equation}
X_{N}=X_{0}(Z_{0}Z_{1}Z_{2} \cdots Z_{N-1})=X_{0}\prod^{N-1}_{n=0}Z_{n}
\end{equation}

where  $X_{0}$ denotes the value of company fundamentals at the initial point. 

\textbf{Assumption 2:} The number, $N$ , of stochastic variables, $Z_{n}$ , differs by company. Furthermore, we assume that   $N$ follows a geometric distribution. That is, 
\begin{equation}
f_{N}(n)=Pr(N=n)=pq^{n-1} \quad for \quad  n=1,2,\cdots
\end{equation}

where $0<p<1$ , and  $q=1-p$.  Thus, the value of company fundamentals,$\bar{X}=X_{0}\prod^{N-1}_{n=0}Z_{n}  $ , is a variable, where $N$  is a random variable following (10) .

Given Assumption 1 and Assumption 2, we have the following Proposition: 

(i) If the probability that variable $Z_{n}$  is greater than one is non-negative, that is, $Pr(Z_{n}>1)>0$  , then the upper tail of the complementary cumulative distribution of $X_{n}$  can be approximated by the following power law distribution: 

\begin{equation}
Pr(\bar{X}>1) \sim cx^{-\alpha}, \quad as \quad x \to \infty   
\end{equation}

where $c$ and  $\alpha$ are positive constants. 

 (ii) If  $\lambda=\frac{p}{q} \to 0$, then $\alpha \to 1$ . 

The proof is given by Reed (2004). For a sketch of the proof, see the Appendix. 

As described above, if there are a few companies that have a large number of factors positively affecting company fundamentals, then company fundamentals follow Zipf's law. Thus, to the degree that the stock market can properly evaluate company fundamentals, company share price will follow the same distribution as company fundamentals. 

\section{Conclusion}
This paper considers the reason why share prices follow Zipf's law. To this end, we investigate the relationship between company fundamentals and share price. We use a database containing financial information for approximately 8,000 globally listed companies and estimate company fundamentals using a panel regression model (two-way fixed effects model).
 
We find that the distribution of company fundamentals follows Zipf's law, and, moreover, that the distribution of fundamentals matches substantially the distribution of actual share prices. From these findings, we conclude that Zipf's law for share prices reflects Zipf's law for company fundamentals. More generally, to the extent that the stock market has the ability to properly evaluate company fundamentals, the distribution of share price reflects company fundamentals. We also find that three common financial indicators - dividends per share, cash flow per share, and book value per share - follow Zipf's law. These findings appear to suggest that Zipf's law for company fundamentals has a robust disposition.
 
We show a simple stochastic model to explain Zipf's law for company fundamentals. However, the question of why the extreme intercompany disparities described by Zipf's law exist in a capitalistic economy remains mysterious and unexplored.

\section*{Appendix}
\subsection*{A: Table of significance values for LMZ for use in Urzua (2000)}
\begin{table}[h]
\begin{center}
\begin{tabular}{cccccccccc} \hline
n           & 10 & 15 & 20 & 25 & 30 & 50 & 100 & 200 & $\infty$  \\ \hline
Level     &     &     &       &     &      &      &       &       &            \\  
 5\%       & 6.19 & 6.14 & 6.09 &6.08& 6.03& 5.98 & 5.98 & 5.99 & 5.99 \\  
 10\%     & 4.38 & 4.41 &4.43 & 4.45 & 4.46 & 4.49 & 4.56 & 4.58 & 4.61 \\ \hline

\end{tabular}
\end{center} 
\end{table}
  
Source: Own Monte Carlo simulation using inversion method, and after 1000,000 replications

\subsection*{B: Derivations of the maximum likelihood estimator (MLE) for the shape parameter of a power law.}

In equation (1) , we described a power law distribution as a complementary cumulative distribution with power-law form. The probability density function for the Pareto distribution\footnote{The Pareto distribution is equivalent to a cumulative distribution with power law form}  is defined as 

\begin{equation}
\numberwithin{equation}{section}
 f(x)=\frac{\alpha k^{\alpha}}{x^{\alpha+1}}    \quad  x \ge k > 0  \tag{B.1}
\end{equation}

where $\alpha$ is the shape parameter and $k$  is the scale parameter corresponding to the minimum value of the distribution.
  
The probability density function, $f(x)$ , is given by the following likelihood function 
\begin{equation}
L=\prod^{n}_{i=1}\frac{\alpha k^{\alpha}}{x^{\alpha+1}_{i} } \tag{B.2}
\end{equation}
Logarithm  $L$ of the likelihood function is written as 

\begin{gather}
\ln L =\ln \prod^{n}_{i=1}\frac{\alpha k^{\alpha}}{x^{\alpha+1}_{i}}   \notag \\
       =\sum^{n}_{i=1}[\ln \alpha+\alpha \ln k-(\alpha+1)\ln x_{i}]  \notag \\
      =n\ln \alpha+n\alpha \ln k-(\alpha+1)\sum^{n}_{i=1}\ln x_{i}  \tag{B.3}
\end{gather}
Setting $\partial L/\partial \alpha=0$  and solving for $\alpha$  , we obtain the following MLE for the shape  parameter. 
\begin{gather}
\frac{n}{\alpha}+n\ln k-\sum^{n}_{i=1}\ln x_{i}=0     \tag{B.4}  \\
\alpha=n[\sum^{n}_{i=1}\ln (\frac{x_{i}}{k})]^{-1}      \tag{B.5}  
\end{gather}
Let $\hat{k}=\min_{i} x_{i} $ ,  then 
\begin{equation}
\hat{\alpha}=n[\sum^{n}_{i=1}\ln (\frac{x_{i}}{x_{\min }})]^{-1}  \tag{B.6}
\end{equation}

\subsection*{C: Sketch of a Reed's (2004) proof of the Proposition}

We derive the Paretian tail behavior for the model. The derivation uses generating functions— the probability generating function (pgf), which for a discrete random variable X with pmf, is defined as 

\begin{equation}
G_{X}(s)=E(s^{X})=\sum f_{x}(x)s^{x}  \tag{C.1}
\end{equation}
and the moment generating function (mgf), which for any random variable X is defined as 
\begin{equation}
M_{X}(s)=E(e^{sX})  \tag{C.2}
\end{equation}
provided the expectations exist. For a random variable, N, with a geometric distribution (10), the pgf is 
\begin{equation}
G_{N}(s)=\frac{ps}{1-qs}  \tag{C.3}
\end{equation}

Now let $\bar{Y}=\log (\bar{X})$  , where $\bar{X}$   is a random variable denoting the fundamentals: 
$(\bar{X}=X_{0}\prod^{N-1}_{i=0}Z_{i})$
 . 
Then 
\begin{equation}
 \bar{Y}=Y_{0}+\sum^{N-1}_{i=0}U_{i}   \tag{C.4}
\end{equation}

where $Y_{0}= \log (X_{0}) $ and $U_{i}=\log (Z_{i})$   for $i=0,1,\cdots, N-1$ . The mgf of  $\bar{Y}$ is
 
\begin{equation}
M_{\bar{Y}}(s)=E(e^{s\bar{Y}})=E(\exp [Y_{0}s+\sum^{N-1}_{i=0}U_{i}s])   \tag{C.5}
\end{equation}

where the expectation is taken with respect to the random variables $Y_{0}, N$  and $U_{0},U_{1}, \cdots, U_{N-1}$ , assumed to be independent. Using conditional expectation, this can be written as
\begin{equation}
M_{\bar{Y}}=M_{0}(s) E(\exp [Y_{0}s+\sum^{N-1}_{i=0}U_{i}s])   \tag{C.6}
\end{equation}

where $M_{0}(s)$   is the mgf of $Y_{0}$ ;  $M_{U}(s)$ is the common mgf of the $U_{i}$   and the expectation is taken with respect to the random variable $N$ . Now,$E([M_{U}(s)]^{N})=G_{N}(M_{U}(s))  $, so that the mgf of $\bar{X}$   can be written 
\begin{equation}
M_{\bar{Y}}=M_{0}(s)\frac{pM_{U}(s)}{1-qM_{U}(s)}   \tag{C.7}
\end{equation}

From standard results, the tail behavior of the pdf of $\bar{Y}$  can be determined from the singularities of its mgf. These occur at the solution (in $s$ ) to $M_{U}(s)=1/q$  . We examine these in the case where
fundamentals are able to increase and decrease:  $Pr(Z_{i}>1)>0$ and $Pr(Z_{i}<1)>0$  so that   $Pr(U_{i}>0)>0$    $Pr(U_{i}<0)>0 $ .

$M_{U}(s) \to \infty \quad  as \quad  s \to \infty$　and $s \to -\infty $. 
From this fact and the convexity of  $M_{U}(s)$, it follows that there are two simple poles of  $M_{\bar{Y}}(s)$, one positive  $\alpha$  and the other negative $-\beta$ . This implies
\begin{equation}
f_{\bar{X}}(s) \sim c_{1}x^{-\alpha-1} (as \quad  x \to \infty)  \quad and \quad   f_{\bar{X}}(s) \sim  \tag{C.8} c_{1}x^{\beta-1} (\quad as \quad  x \to 0)
\end{equation}

The asymptotic behavior of the cdf  $F_{\bar{X}}(x)$  or the complementary cdf, $1-F_{\bar{X}}(x)$ , follows from integration. 

We introduce the quantity $\lambda=p/q$  and the quantity $\theta=1-\log (q)$  which are related as $\theta=-\log (1+\lambda)$  , an increasing function. It follows that $\beta$  increases with $\lambda$  . In the limit as $\lambda \to 0, \quad \beta \to 0$ , the complementary cdf, $1-F_{\bar{X}}(x)$  , tends to follow the power law distribution with the exponent  $1+\beta$ equal to unity in the right tail.

\begin{figure}[h]
\includegraphics[width=70mm]{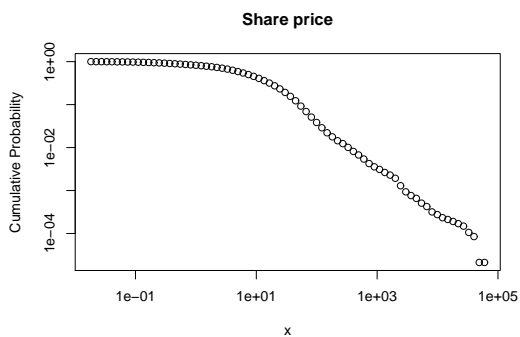}\\
\caption{Zipf's law for share price: The complementary cumulative distribution of the share prices. (log-log plot). The 47,161 share price data is pooled from 7,796 companies for the period 2004-2013. The power-law exponent estimated by MLE (maximum likelihood estimator) method is 1.003.} 
\end{figure}

\begin{figure}[h]
\includegraphics[width=70mm]{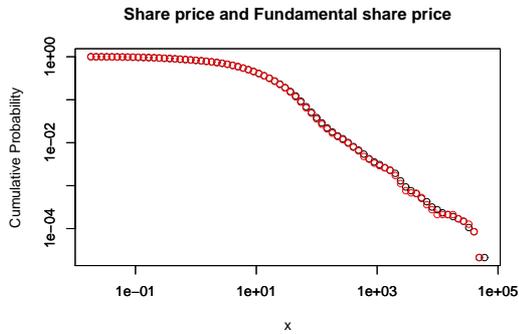}\\  
\caption{The complementary cumulative distribution of actual share price and fundamental (log-log plot). black: actual share price and red: fundamentals The complementary cumulative distribution of fundamentals coincides statistically with that of actual share price } 
\end{figure}

\begin{figure}
\includegraphics[width=70mm]{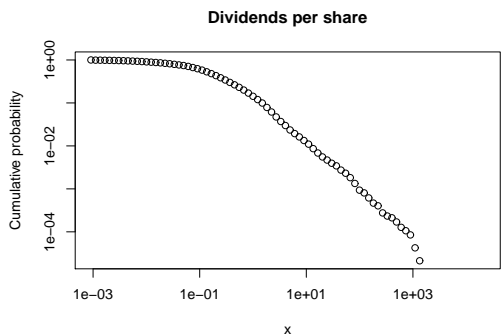}\\
\caption{Zipf's law for the dividends per share. The complementary cumulative distribution of the dividends per share (log-log plot). The 47,161 dividends per share is pooled from 7,796 companies for the period 2004-2013. The power-law exponent estimated by MLE method is 1.015}
\end{figure}
\begin{figure}[h]
\includegraphics[width=70mm]{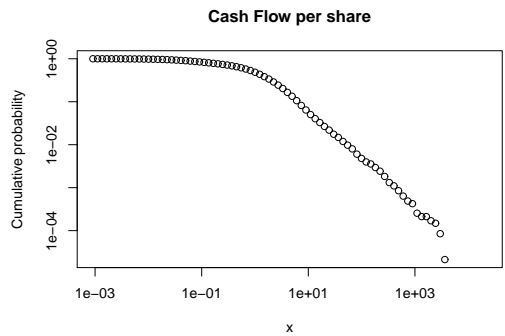}\\  
\caption{Zipf's law for cash flow per share. The complementary cumulative distribution of the cash flow per share (log-log plot). The 47,161 cash flow per share is pooled from 7,796 companies for the period 2004-2013. The power-law exponent estimated by MLE method is 1.051 }
\end{figure}
\begin{figure}[h]
\includegraphics[width=70mm]{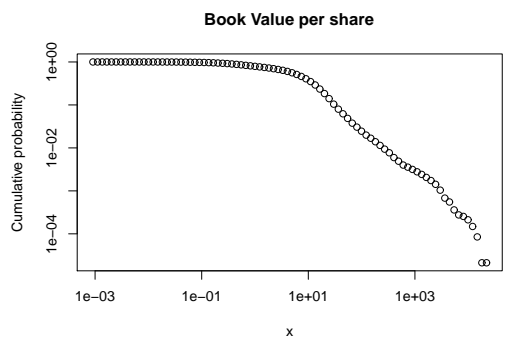}\\
\caption{Zipf's law for cash flow per share. The complementary cumulative distribution of the book value per share (log-log plot). The 47,161 book value per share is pooled from 7,796 companies for the period 2004-2013. The power-law exponent estimated by MLE method is 0.955 }
\end{figure}

\end{document}